# Hybridization of Graphene-Gold Plasmons for Active Control of Mid-Infrared Radiation

Matthew Feinstein, Euclides Almeida

*Department of Physics, Queens College, City University of New York, Flushing, NY 11367, United States of America*

*The Graduate Center of the City University of New York, New York, NY 10016, United States of America*

*Author's email address: euclides.almeida@qc.cuny.edu

**Abstract**

Many critical applications in environmental and biological sensing, standoff detection, and astronomy rely on devices that operate in the mid-infrared range. Unfortunately, current mid-infrared devices are costly and have limited tunability. Nanostructured graphene has been proposed for active mid-infrared devices via excitation of tunable surface plasmons, but typically present low efficiencies due to weak coupling with free-space radiation and plasmon damping. We present an alternative approach, in which graphene plasmons couple with gold localized plasmons, creating hybridized modes and enhancing the coupling efficiency. We demonstrate a metasurface in which hybrid plasmons are excited with transmission modulation rates of 17% under moderate doping (0.35 eV). We also evaluate the metasurface as a mid-infrared modulator, measuring switching speeds of up to 16 kHz. Finally, we propose a scheme in which we can excite strongly coupled gold-graphene gap plasmons in the thermal radiation range, with applications to nonlinear optics, slow light, and sensing.

**Introduction**

Devices operating in the mid-infrared region of the electromagnetic spectrum find significant applications in imaging, sensing, communication, and astronomy due to the presence of unique molecular "spectral fingerprints" and atmospheric transparency windows between 3-5 µm and 8-12 µm. However, the development of mid-infrared technology has been hindered by the lack of materials with a small enough bandgap and the requirement to operate at lower temperatures. Graphene, with no bandgap in a charge-neutral state, can absorb 2.3% of light across the THz and mid-infrared range, as determined by the fine-structure constant [1]. Although this is an impressive value for a single-atom-thick material, for applications such as metasurface modulators this value is insufficient [2,3]. Nevertheless, nanostructured graphene can couple to light through localized surface plasmons and enhance light absorption and scattering [4-8]. Moreover, graphene plasmon resonances can be tuned across a significant part of the mid-infrared through charge injection. The coupling, however, is inefficient due to wavevector mismatch, the thinness of graphene, plasmon damping, and interaction with the substrate [5,6]. In particular, direct patterning of the graphene introduces defects at the edges which render the edges electrically inactive [6]. Measured mid-infrared extinction rates in graphene nanostructures have been consistently smaller than 10% [4-6,9-12], with a notable exception showing an extinction of 40% at high (0.8 eV) doping in a top-gated scheme with a high-capacitance ion-gel as the gate dielectric [11].

More recently, alternative schemes have been proposed to enhance absorption, making use of gold antennas or slots to interact with graphene plasmons [13-16]. Kim *et al.* showed near perfect absorption in a graphene nanoribbons metasurface by engineering mirror charges on a low permittivity substrate and demonstrated a mid-infrared modulator operating in reflectance [15]. Zeng *et al.* demonstrated a metasurface mid-infrared modulator with reflection modulation depth of up to 90% and a spatial light modulator device [16]. These other exciting results have showed the viability of using graphene as a platform for active mid-infrared optical elements and devices.

Here we propose an alternative approach, based on hybridization of gold-graphene plasmons, in which their strong near-field interaction can create tunable and sharp plasmonic resonances across the mid-infrared range. In this scheme, the graphene itself does not need to be etched (avoiding the creation of inert regions of graphene), and the gold rods enhance the coupling with light. We demonstrate a metasurface in which a hybrid, graphene-like plasmon resonance can be tuned through the application of an external voltage, allowing for high differential transmission rates of 20% under moderate doping conditions. We also explore the potential application of such a device as an electro-optic modulator by applying an oscillating voltage source to the gate and measuring the optical modulation. Finally, we propose a design for a metasurface operating in the thermal radiation range showing strong coupling between graphene and gold plasmons, with potential applications in sensing and nonlinear optics.

**Hybridization model**

The concept of hybridization in plasmonics was introduced by Prodan et al. to explain plasmon resonances in metallic nanostructures [17]. Originally applied to metallic nanoshells, it has been used to explain and design plasmon resonances in a myriad complex structures [18-20], including in nanostructured graphene [10,21]. This powerful concept borrowed ideas from hybridization in chemistry to provide an intuitive picture to explain the plasmon resonances and visualize charge distribution. In our case, the metasurface design, depicted in Fig. 1(a), will be based on hybridization between graphene and gold plasmons. It consists of gold nanorods deposited on a graphene surface in a periodic configuration. Mid-infrared light excites localized surface plasmons (LSPs) in the gold nanorods, which in turn excite graphene LSPs confined in the gaps between the tips of the gold rods [22,23]. Due to their proximity, they strongly couple, forming hybrid modes. We can control the detuning between graphene and gold LSP frequencies through their geometry and gate voltage. We will consider two cases separately: when these two resonances are far apart, and when they match. Figure 1(b) shows the concept for hybridization in our design. Graphene and gold plasmons will couple through near-field, giving rise to two new hybrid modes, depending on their charge density configuration. In the *bonding* (symmetric) mode, the charge density will be concentrated mostly on graphene, while in the *antibonding* (antisymmetric) mode the charges will be distributed on both graphene and gold rods. The picture is equivalent to that of a coupled harmonic oscillator, but the hybridization concept provides a microscopic description of the coupling origin. The resonances will be calculated from electromagnetic simulations, including a model for the each material permittivity or surface conductivity (see supplementary information).

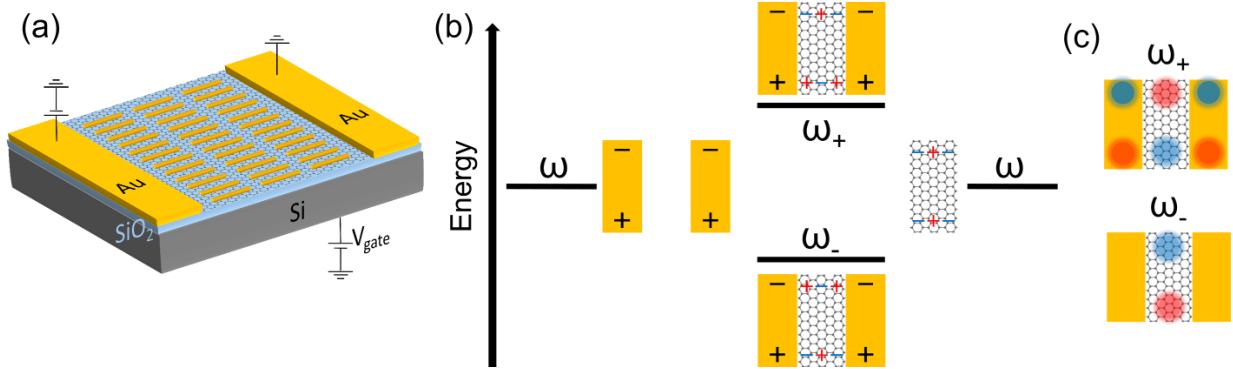

**Figure 1. Hybridization between graphene and gold plasmons. a**) Design of the hybrid gold-graphene metasurface device. The thickness of the SiO$_2$ dielectric spacer is 90 nm. The graphene layer is placed between the gold rods and the SiO$_2$ spacer. b) Microscopic depiction of the hybridization phenomenon. The incident light is polarized along the long (vertical) axis of the gold rod. The near-field around the gold rods couples to dark plasmons in graphene along the horizontal direction. The gold and graphene plasmons hybridize, forming two new modes, whose energies $\hbar\omega_-$ and $\hbar\omega_+$ depend on the charge distribution on graphene. C) Net charge density of antisymmetric ($\omega_+$) and symmetric ($\omega_-$) modes. The red and blue clouds represent opposite charges.

## Results

*Detuned resonances*

We begin by considering the scenario where the LSPs of both gold and graphene are detuned. The design is presented in Fig. 2(a) and the calculated spectra for varying graphene Fermi level are presented in 2(b). When the Fermi level is set to 0, the spectrum displays only a broad resonance of the gold rod, peaked at 4 μm and 9.6 μm. The additional features in the spectrum will be explained in the experimental section later.

By adjusting the Fermi level of graphene, resonances centered around 7.5 µm and in the 11-13 µm range can be observed. The original peak around 4 µm is also shifted by the coupling to the graphene LSP, though this effect is more limited. While the modes are hybridized, the graphene and gold LSPs generally retain their original character within the range of chemical potentials that we can access. Therefore, for detuned resonances, we will refer to these hybrid modes as *graphene-like* and *gold-like*, respectively. In Fig. 2(c), the calculated charge distribution on this metasurface at 7.5 µm confirm that this resonance is *graphene-like*.

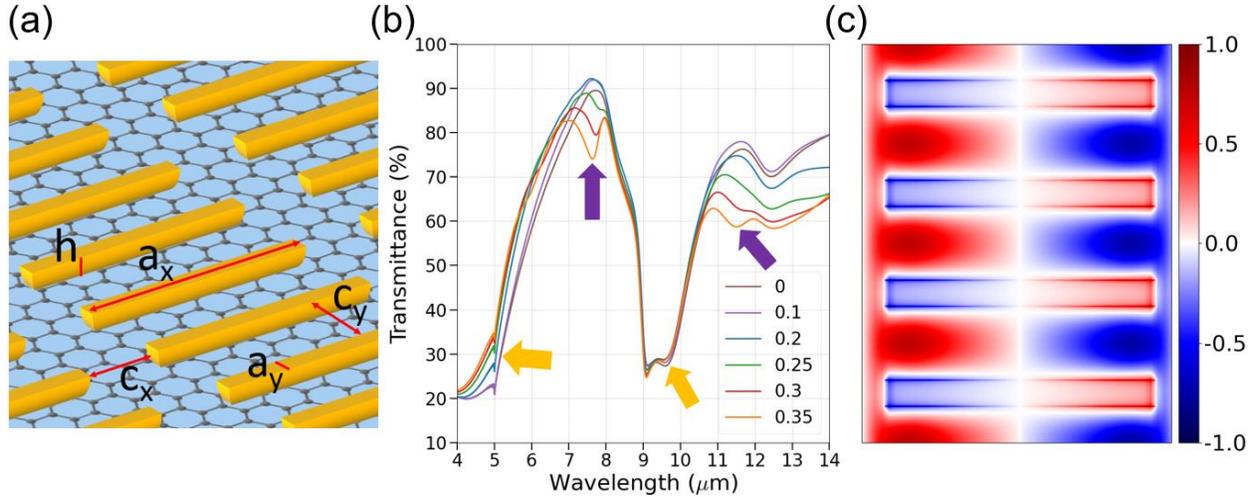

**Figure 2. Design of the hybridized gold-graphene metasurface for detuned resonances.** a) Parameters of the metasurface: $a_x$ = 1210 nm, $a_y$ = 70 nm, $c_x$ = 270 nm, $c_y$ = 150 nm. The height *h* of the gold rods is 25 nm. Light is polarized along the x-axis. b) Calculated transmittance of the metasurface across the mid-infrared regime for various values of graphene's Fermi level in eV. The gold and purple arrows depict "gold-like" and "graphene-like" resonances, respectively. The damping parameter of Graphene is 0.02 eV c) Charge density map (arbitrary units) calculated at λ = 7.5 µm. The charges are mostly concentrated on the graphene area, and the resonance character is essentially that of graphene plasmons.

Figure 3(a) shows an SEM image of the fabricated metasurface, using e-beam lithography on a CVD-grown graphene field-effect transistor device [24] (see methods section). The total area of the metasurface is 250x250 µm². The spectral response of the fabricated device was measured using a Fourier Transform Infrared Spectrometer (FTIR) and is shown in Fig. 3(c). The gate voltage was adjusted to tune the graphene chemical potentials from 0 to 0.35 eV. Figure 3(d) shows the differential extinction, defined here as $\Delta T(E_F) = 1 - T(E_F)/T_{CNP}$, where $T(E_F)$ and $T_{CNP}$ are transmittances at the graphene's Fermi level $E_F$ and at charge neutrality point (CNP, $E_F = 0$), respectively. This metric is useful for comparing the performance of our graphene metasurface at a given doping level to that of graphene at the CNP. The FTIR data shows a strong differential extinction of 17% at 11.5 µm, which corresponds to an atmospheric window, and 11% at 7.5 µm. These two resonances correspond to a *graphene-like* plasmon. Some negative differential transmission can be seen where the *gold-like* LSP shifts, which results in higher transmission. The gold rod resonances peak around 4 and 9.6 µm, and coupling with the silica phonons [5,6] in the 9.6 µm region results in a surface enhanced infrared absorption (SEIRA) effect [25,26]. A second peak around 12.8 µm is sensitive to the tuning of the graphene chemical potential, which is the result of splitting of the resonance due to the silica surface optical phonon around 12.4 µm [27]. The linewidths of the graphene-like LSPs are narrower than the gold LSPs due to the lower losses in graphene.

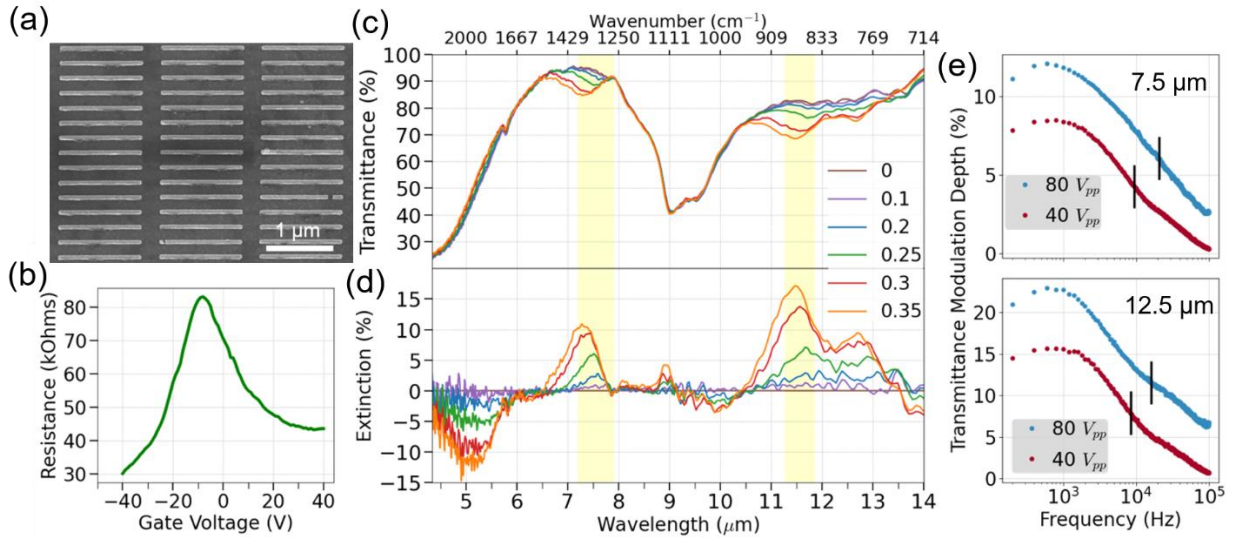

**Figure 3. Characterization of the fabricated metasurface.** a) Scanning electron microscopy of a section of the device. b) Transfer curve of the field-effect transistor device. c) Measured transmittance as a function of graphene's Fermi level, given in eV. d) Calculated differential extinction. e) Frequency response characterization of the mid-infrared modulator for sinusoidal gate voltages of 80 V (blue) and 40 V (red) peak-to-peak. The modulation experiment was conducted with an optical bandpass filter at 7.5 µm (top) and 11.5 µm (bottom), which correspond to the yellow bands in (c) and (d). Cutoff frequencies (-3 dB, 50% compared to max modulation) are marked with vertical black lines.

Additionally, we conducted an electro-optical modulation experiment by applying an alternating voltage source to the gate over a range of modulation frequencies while measuring the transmission modulation at a select wavelength. The wavelength was selected using a mid-infrared band-pass filter, and the measurements were taken with a lock-in amplifier. The transmission modulation data (Fig. 3(e)) shows strong modulation depths of up to 23% of 11.5 µm light and remains within 3 dB (50%) of the maximum modulation depth up to around 16 kHz. This demonstrates the device's potential as an electro-optic modulator or as a laser noise eater. The sensing application is especially promising given that the device is operable in an atmospheric window while in contact with the ambient atmosphere and at room temperature. These results were achieved using CVD graphene on standard substrates (silica on silicon) that was processed through several rounds of lithographic processes, and thus represent a reasonable picture of such a device produced at scale.

The extinction spectrum of the device is significantly affected by the type of substrate used as well as the mobility of graphene. Our experiments showed that using a silica substrate, the device produced multiple resonances due to the dispersion of the refractive index [28], but the tunability of the spectrum was limited. To address this limitation, we conducted numerical simulations for $Al_2O_3$, which are presented in Supplementary Figure 1. These simulations showed that by using $Al_2O_3$ substrate, we could tune the graphene resonances from 9 to 6.5 µm, providing greater spectral tunability. Additionally, we found that even with low mobility graphene, we were still able to achieve a significant response

*In-tune resonances*

We now focus our attention to the case where gold and graphene plasmonic resonances have the same frequency. Gold and graphene are materials whose optimal plasmonic resonances fall in different parts of the electromagnetic spectrum and are considered to be complementary [29,30]. While gold presents outstanding performance in the visible and near-infrared ranges, in the mid-infrared its optical extinction coefficient is too high and the resonances are broad. Graphene plasmons, on the other hand, can be fine-tuned to the mid-infrared up to around 6.4 µm, beyond which they decay due to optical phonons [6]. Here, we will make use of gap plasmons to tune the gold rod resonance to the 8-10 µm window. The concept of gap plasmons has been thoroughly studied and documented [31-34], and the mechanism for shifting the resonance to the mid-infrared is similar to that of hybridization [31].

The design is shown in Figure 4(a). The gold rods are separated from a gold plane by a thin 15 nm dielectric ($Al_2O_3$) layer and graphene. When coupled to light, the rods induces mirror charges on the metal plane beneath it, and the plane and the rods couple. As a result, the bonding mode arising from this interaction falls in the mid-infrared around 9.1 $\mu m$, as seen in the spectrum in Fig. 4(c) for a graphene Fermi level of 0 eV. By increasing the Fermi level of graphene, graphene plasmons are excited and hybridization between graphene-gap and gold-gap plasmons takes place. We can observe the new two hybridized modes, and the charge density distribution and near-field images shown in Figs. 4(e) and 4(f) is in line with the hybridization picture in Fig. 1. By using doping as the detuning parameter, we observe an anti-crossing behavior, and at the point of minimal detuning ($E_F = 0.18$ eV), the separation between the hybrid modes is $2g = 23.0$ meV.

We compare this coupling value to the widths of the resonances of the hybrid modes $y_+ = 24.9$ meV and $y_- = 8.3$ meV. In this design, the coupling rate $2g > (y_+ + y_-)/2$, indicating the system is in the strong coupling regime. In this regime, the two modes exchange energy at a rate faster than their dissipation rate, resulting in longer-lived plasmons and narrowed resonances. This is evident from the comparison of the widths of gold gap plasmons at $E_F = 0$ (26.6 meV) and hybridized resonances.

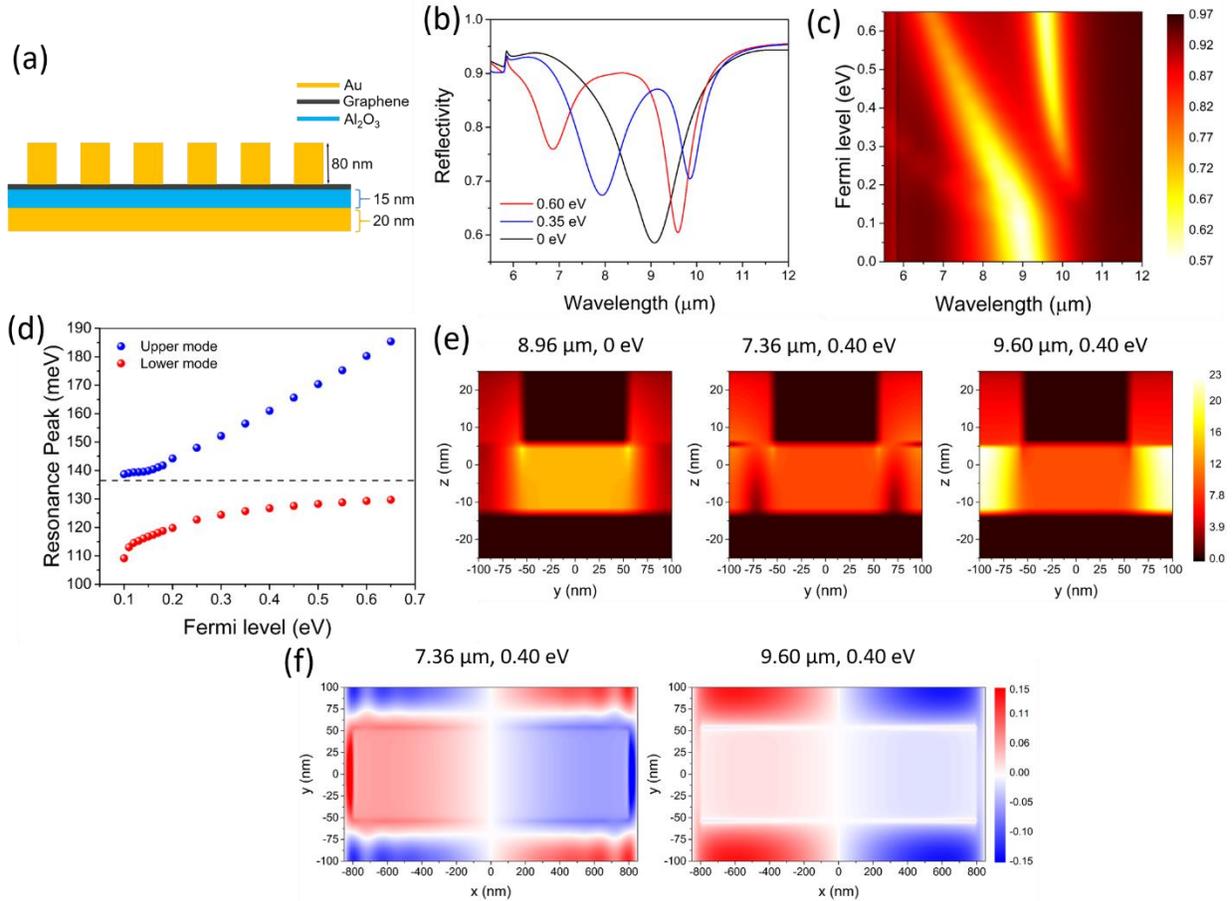

**Figure 4. Design of hybridized gold-graphene metasurfaces for in-tune resonances.** a) Illustration of the metasurface's cross-section. The parameters of the periodic array (top view shown in Fig. 2(a)) are $a_x = 1600$ nm, $a_y = 110$ nm, $c_x = 90$ nm, $c_y = 100$ nm. b) Calculated reflectivity of the metasurface for three different values of graphene's Fermi level, showing tunable symmetric and antisymmetric modes. The damping parameter of graphene is set to 13 meV c) Reflectivity of the metasurface calculated for a larger range of graphene's Fermi level. d) Resonance peak for several Fermi levels. The minimum separation between the hybrid modes is at 0.18 eV, and corresponds to $\Delta = 23$ meV. e) Calculated near-field distribution of the gap-plasmons for $E_F = 0$ eV and $\lambda = 8.96$ μm (left), $E_F = 0.40$ eV and $\lambda = 7.36$ μm (center), and $E_F = 0.40$ eV and $\lambda = 9.60$ μm (right). The field was calculated at the position x=700nm. f) Calculated charge density map (arbitrary units) for $E_F = 0.40$ eV and $\lambda = 7.36$ μm (left), $E_F = 0.40$ eV and $\lambda = 9.60$ μm (right).

In our assumptions above, we slightly underestimate the coupling strength between gold and graphene plasmons. This is because tuning graphene's Fermi level also changes the damping rate of graphene plasmons. Specifically, at lower doping the interband damping is more pronounced, which inhibits the coupling strength. the coupled harmonic oscillator model, the frequency of the hybrid (normal) modes, are given by [35,36]

$$\omega_\pm = \omega_0 - i\left(\frac{\gamma_{Au}}{2} + \frac{\gamma_{gr}}{2}\right) \pm \frac{1}{2}\sqrt{\Omega^2 - \left(\frac{\gamma_{Au}}{2} - \frac{\gamma_{gr}}{2}\right)^2} \quad (1)$$

where in our case, $\omega_0$ is the resonance frequency at $E_F = 0$ eV, $\gamma_{Au}$ and $\gamma_{gr}$ are the damping rates of gold and graphene plasmons respectively, and $\Omega$ is the gold-graphene plasmons interaction potential ($\Omega = 2g$ for $\gamma_{Au} = \gamma_{gr}$). Therefore, the frequency of the hybrid modes should be equidistant to $\omega_0$. This occurs at $E_F = 0.27$ eV, in which $\omega_+ = 149.7$ meV, $\omega_- = 123.5$ meV, and the coupling rate $2g = 26.2$ meV.

Finally, as we increase the Fermi level, the charge density distribution is still concentrated on graphene for the bonding mode (*graphene-like*, right panel of Fig. 4(f)), as expected by the hybridization picture provided in Fig. 1(c). For the antibonding mode however, the charge density is distributed on gold and graphene (left panel of Fig. 4(f)), assuming a truly hybrid character. In the left charge density map of Fig. 4f, we can observe a spatial modulation of the charge density. This is caused by propagating graphene plasmons, which are launched by the gold rods along the x-axis, and interfere with the localized plasmons between the rods (along the y-axis).

**Discussion and summary**

We have presented a simple design that creates a hybridization between graphene and gold plasmons, offering a microscopic view of their interaction and enabling the creation of more sophisticated structures in the future to improve field localization. Using a commercially available, large-area CVD-grown graphene field effect transistor, we have demonstrated narrow resonances in the mid-infrared region, achieving up to 17% transmission modulation with only a modest 0.35 eV chemical doping. Our device also showed dual band modulation of 20% at a (3dB) rate of up to 16 kHz, with potential improvements in switching speed by reducing the device area and consequently the device's sheet resistivity.

The tunable, narrow resonances of our design make it highly desirable for mid-infrared metasurfaces. A metasurface is composed of an array of subwavelength particles on a surface that can couple with incident light [31,37-41]. The geometry of each particle in the array can be designed to alter the amplitude, phase, and polarization of the incident light at a subwavelength scale, which can effect such functionalities as lensing

[42,43], beam-steering [41,44], holography [45,46], and nonlinearity [47-50]. Our metal-on-mirror design with matched resonances offers two distinct regions where the amplitude and phase of mid-infrared radiation can be controlled. Moreover, the sharp dip at the center of the spectrum resembles an electromagnetic-induced transparency dip that could be used to observe electrically tunable "slow light" [51] in the mid-infrared range.

The device's extinction rate can be improved by optimizing parameters such as the dielectric gate material, dielectric thicknesses, array dimensions, and index matching between the top and bottom surfaces. Calculations for a metasurface using $Al_2O_3$ as the gate dielectric are presented in the Supplementary Information and showed enhanced extinction rates. When searching for the best parameters, either rational (from hybridization theory) or inverse design methods [52] can be applied, or even a combination of both.

The fact that our design is exposed to air may greatly benefit applications such as biosensing [53] and gas detection [54]. Moreover, the multiple resonances can enhance coherent anti-Stokes Raman signals. Finally, one can visualize applications such as tunable mid-infrared band-pass filters, active control of thermal radiation and cooling.

**Supplementary information**

Methods, device design and fabrication, optical and electrical characterization, and calculations of detuned metasurfaces using $Al_2O_3$ substrate can be found in the supplementary information.

**Acknowledgements**

This work was partially conducted at the nanofabrication facilities of the Advanced Science Research Center at the graduate center of CUNY. The authors acknowledge useful discussions with Dr. Milan Begliarbekov regarding sample fabrication.

**Data availability statement**

Data supporting this work are available upon request.

**Conflict of interest**

The authors declare no conflict of interest.

References

1. Nair, R. R.; Blake, P.; Grigorenko, A. N.; Novoselov, K. S.; Booth, T. J.; Stauber, T.; Peres, N. M. R.; Geim, A. K. Fine Structure Constant Defines Visual Transparency of Graphene. *Science* **2008**, *320*, 1308.

2. Yu, S.; Wu, X.; Wang, Y.; Guo, X.; Tong, L. 2D Materials for Optical Modulation: Challenges and Opportunities. *Advanced materials (Weinheim)* **2017**, *29*, 1606128-n/a.

3. Sun, Z.; Martinez, A.; Wang, F. Optical Modulators with 2D Layered Materials. *Nature photonics* **2016**, *10*, 227-238.

4. de Abajo, F. J. G. Graphene plasmonics: Challenges and opportunities. *ACS photonics* **2014**, *1*, 135-152.

5. Atwater, H. A.; Lopez, J. J.; Sherrott, M.; Jang, M. S.; Brar, V. W. Highly Confined Tunable Mid-Infrared Plasmonics in Graphene Nanoresonators. **2013**.

6. Yan, H.; Low, T.; Zhu, W.; Wu, Y.; Freitag, M.; Li, X.; Guinea, F.; Avouris, P.; Xia, F. Damping pathways of mid-infrared plasmons in graphene nanostructures. *NATURE PHOTONICS* **2013**, *7*, 394-399.

7. Low, T.; Avouris, P. Graphene Plasmonics for Terahertz to Mid-Infrared Applications. *ACS nano* **2014**, *8*, 1086-1101.

8. Brar, V. W.; Sherrott, M. C.; Jang, M. S.; Kim, S.; Kim, L.; Choi, M.; Sweatlock, L. A.; Atwater, H. A. Electronic modulation of infrared radiation in graphene plasmonic resonators. *Nature communications* **2015**, *6*, 7032.

9. Alcaraz Iranzo, D.; Nanot, S.; Dias, E. J. C.; Epstein, I.; Peng, C.; Efetov, D. K.; Lundeberg, M. B.; Parret, R.; Osmond, J.; Hong, J.; Kong, J.; Englund, D. R.; Peres, N. M. R.; Koppens, F. H. L. Probing the ultimate plasmon confinement limits with a van der Waals heterostructure. *Science (American Association for the Advancement of Science)* **2018**, *360*, 291-295.

10. Fang, Z.; Thongrattanasiri, S.; Schlather, A.; Liu, Z.; Ma, L.; Wang, Y.; Ajayan, P. M.; Nordlander, P.; Halas, N. J.; García de Abajo, F. J. Gated tunability and hybridization of localized plasmons in nanostructured Graphene. *ACS NANO* **2013**, *7*, 2388-2395.

11. Fang, Z.; Wang, Y.; Schlather, A. E.; Liu, Z.; Ajayan, P. M.; García de Abajo, F. J.; Nordlander, P.; Zhu, X.; Halas, N. J. Active Tunable Absorption Enhancement with Graphene Nanodisk Arrays. *Nano letters* **2014**, *14*, 299-304.

12. Guo, Q.; Yu, R.; Li, C.; Yuan, S.; Deng, B.; García de Abajo, F. J.; Xia, F. Efficient electrical detection of mid-infrared graphene plasmons at room temperature. *Nature materials* **2018**, *17*, 986-992.

13. Kim, S.; Jang, M. S.; Brar, V. W.; Tolstova, Y.; Mauser, K. W.; Atwater, H. A. Electronically tunable extraordinary optical transmission in graphene plasmonic ribbons coupled to subwavelength metallic slit arrays. *Nature communications* **2016**, *7*, 12323.

14. Zhang, J.; Guo, C.; Liu, K.; Zhu, Z.; Ye, W.; Yuan, X.; Qin, S. Coherent perfect absorption and transparency in a nanostructured graphene film. *Optics express* **2014**, *22*, 12524-12532.

15. Kim, S.; Jang, M. S.; Brar, V. W.; Mauser, K. W.; Kim, L.; Atwater, H. A. Electronically Tunable Perfect Absorption in Graphene. *Nano letters* **2018**, *18*, 971-979.

16. Zeng, B.; Huang, Z.; Singh, A.; Yao, Y.; Azad, A. K.; Mohite, A. D.; Taylor, A. J.; Smith, D. R.; Chen, H. Hybrid graphene metasurfaces for high-speed mid-infrared light modulation and single-pixel imaging. *Light, science & applications* **2018**, *7*, 51-8.

17. Prodan, E.; Radloff, C.; Halas, N. J.; Nordlander, P. A Hybridization Model for the Plasmon Response of Complex Nanostructures. *Science* **2003**, *302*, 419-422.

18. Tserkezis, C.; Esteban, R.; Sigle, D. O.; Mertens, J.; Herrmann, L. O.; Baumberg, J. J.; Aizpurua, J. Hybridization of plasmonic antenna and cavity modes: Extreme optics of nanoparticle-on-mirror nanogaps. *Physical review. A, Atomic, molecular, and optical physics* **2015**, *92*.

19. Halas, N. J.; Lal, S.; Chang, W.; Link, S.; Nordlander, P. Plasmons in Strongly Coupled Metallic Nanostructures. *Chemical reviews* **2011**, *111*, 3913-3961.

20. Nordlander, P.; Oubre, C.; Prodan, E.; Li, K.; Stockman, M. I. Plasmon Hybridization in Nanoparticle Dimers. *Nano letters* **2004**, *4*, 899-903.

21. Mou, N.; Sun, S.; Dong, H.; Dong, S.; He, Q.; Zhou, L.; Zhang, L. Hybridization-induced broadband terahertz wave absorption with graphene metasurfaces. *Optics express* **2018**, *26*, 11728-11736.

22. FEI, Z.; RODIN, A. S.; FOGLER, M. M.; CASTRO, A. H.; LAU, C. N.; KEILMANN, F.; BASOV, D. N.; ANDREEV, G. O.; BAO, W.; MCLEOD, A. S.; WAGNER, M.; ZHANG, L. M.; ZHAO, Z.; THIEMENS, M.; DOMINGUEZ, G. Gate-tuning of graphene plasmons revealed by infrared nano-imaging. *Nature (London)* **2012**, *487*, 82-85.

23. Alonso-González, P.; Nikitin, A. Y.; Golmar, F.; Centeno, A.; Pesquera, A.; Vélez, S.; Chen, J.; Navickaite, G.; Koppens, F.; Zurutuza, A.; Casanova, F.; Hueso, L. E.; Hillenbrand, R. Controlling graphene plasmons with resonant metal antennas and spatial conductivity patterns. *Science (American Association for the Advancement of Science)* **2014**, *344*, 1369-1373.

24. Novoselov, K. S.; Geim, A. K.; Morozov, S. V.; Jiang, D.; Zhang, Y.; Dubonos, S. V.; Grigorieva, I. V.; Firsov, A. A. Electric Field Effect in Atomically Thin Carbon Films. *Science* **2004**, *306*, 666-669.

25. Neubrech, F.; Huck, C.; Weber, K.; Pucci, A.; Giessen, H. Surface-Enhanced Infrared Spectroscopy Using Resonant Nanoantennas. *Chemical reviews* **2017**, *117*, 5110-5145.

26. Kundu, J.; Le, F.; Nordlander, P.; Halas, N. J. Surface enhanced infrared absorption (SEIRA) spectroscopy on nanoshell aggregate substrates. *Chemical physics letters* **2008**, *452*, 115-119.

27. Gunde, M. K. Vibrational modes in amorphous silicon dioxide. *Physica. B, Condensed matter* **2000**, *292*, 286-295.

28. Palik, E. D. *Handbook of Optical Constants of Solids;* Elsevier: San Diego, 1998; .

29. Pile, D. Graphene versus metal plasmons. *Nature photonics* **2013**, *7*, 420.

30. Buljan, H.; Jablan, M.; Soljačić, M. Damping of plasmons in graphene. *Nature photonics* **2013**, *7*, 346-348.

31. Tserkezis, C.; Esteban, R.; Sigle, D. O.; Mertens, J.; Herrmann, L. O.; Baumberg, J. J.; Aizpurua, J. Hybridization of plasmonic antenna and cavity modes: Extreme optics of nanoparticle-on-mirror nanogaps. *Physical review. A, Atomic, molecular, and optical physics* **2015**, *92*.

32. Neutens, P.; Van Dorpe, P.; De Vlaminck, I.; Lagae, L.; Borghs, G. Electrical detection of confined gap plasmons in metal-insulator-metal waveguides. *Nature photonics* **2009**, *3*, 283-286.

33. Ding, F.; Yang, Y.; Deshpande, R. A.; Bozhevolnyi, S. I. A review of gap-surface plasmon metasurfaces: fundamentals and applications. *Nanophotonics* **2018**, *7*, 1129-1156.

34. Kim, J.; Lee, C.; Lee, Y.; Lee, J.; Park, S.; Park, S.; Nam, J. Synthesis, Assembly, Optical Properties, and Sensing Applications of Plasmonic Gap Nanostructures (Adv. Mater. 46/2021). *Advanced materials (Weinheim)* **2021**, *33*, 2170360-n/a.

35. Törmä, P.; Barnes, W. L. Strong coupling between surface plasmon polaritons and emitters: a review. *RoPP* **2015**, *78*, 013901.

36. Savona, V.; Andreani, L. C.; Schwendimann, P.; Quattropani, A. Quantum well excitons in semiconductor microcavities: Unified treatment of weak and strong coupling regimes. *Solid state communications* **1995**, *93*, 733-739.

37. Yu, N.; Genevet, P.; Kats, M. A.; Aieta, F.; Tetienne, J.; Capasso, F.; Gaburro, Z. Light Propagation with Phase Discontinuities: Generalized Laws of Reflection and Refraction. *SCIENCE* **2011**, *334*, 333-337.

38. Yu, N.; Capasso, F. Flat optics with designer metasurfaces. *Nature materials* **2014**, *13*, 139-150.

39. Lin, D.; Fan, P.; Hasman, E.; Brongersma, M. L. Dielectric gradient metasurface optical elements. *Science (American Association for the Advancement of Science)* **2014**, *345*, 298-302.

40. Bomzon, Z.; Kleiner, V.; Hasman, E. Pancharatnam–Berry phase in space-variant polarization-state manipulations with subwavelength gratings. *Optics letters* **2001**, *26*, 1424-1426.

41. Ni, X.; Emani, N. K.; Kildishev, A. V.; Boltasseva, A.; Shalaev, V. M. Broadband Light Bending with Plasmonic Nanoantennas. *Science (American Association for the Advancement of Science)* **2012**, *335*, 427.

42. Khorasaninejad, M.; Capasso, F. Metalenses. *Science (American Association for the Advancement of Science)* **2017**, *358*, 1146.

43. Avayu, O.; Almeida, E.; Prior, Y.; Ellenbogen, T. Composite functional metasurfaces for multispectral achromatic optics. *Nature Communications* **2017**, *8*, 14992.

44. Arbabi, A.; Horie, Y.; Bagheri, M.; Faraon, A. Dielectric metasurfaces for complete control of phase and polarization with subwavelength spatial resolution and high transmission. *Nature nanotechnology* **2015**, *10*, 937-943.

45. Larouche, S.; Tsai, Y.; Tyler, T.; Jokerst, N. M.; Smith, D. R. Infrared Metamaterial Phase Holograms. *Nature materials* **2012**, *11*, 450-454.

46. Ni, X.; Kildishev, A. V.; Shalaev, V. M. Metasurface holograms for visible light. *Nature communications* **2013**, *4*, 2807.

47. Li, G.; Zhang, S.; Zentgraf, T. Nonlinear photonic metasurfaces. *Nature Reviews Materials* **2017**, *2*, 17010.

48. Segal, N.; Keren-Zur, S.; Hendler, N.; Ellenbogen, T. Controlling light with metamaterial-based nonlinear photonic crystals. *Nature photonics* **2015**, *9*, 180-184.

49. Almeida, E.; Bitton, O.; Prior, Y. Nonlinear metamaterials for holography. *Nature Communications* **2016**, *7*, 12533.

50. Almeida, E.; Shalem, G.; Prior, Y. Subwavelength nonlinear phase control and anomalous phase matching in plasmonic metasurfaces. *Nature Communications* **2016**, *7*, 10367.

51. Zhang, S.; Genov, D. A.; Wang, Y.; Liu, M.; Zhang, X. Plasmon-Induced Transparency in Metamaterials. *Physical review letters* **2008**, *101*, 047401.

52. Molesky, S.; Lin, Z.; Piggott, A. Y.; Jin, W.; Vucković, J.; Rodriguez, A. W. Inverse design in nanophotonics. *Nature photonics* **2018**, *12*, 659-670.

53. Rodrigo, D.; Limaj, O.; Janner, D.; Etezadi, D.; de Abajo, F. J. G.; Pruneri, V.; Altug, H. Mid-infrared plasmonic biosensing with graphene. *Science* **2015**, *349*, 165-168.

54. Bareza, N. J.; Gopalan, K. K.; Alani, R.; Paulillo, B.; Pruneri, V. Mid-infrared Gas Sensing Using Graphene Plasmons Tuned by Reversible Chemical Doping. *ACS photonics* **2020**, *7*, 879-884.

# Supplementary information for

# Hybridization of Graphene-Gold Plasmons for Active Control of Mid-Infrared Radiation

Matthew Feinstein, Euclides Almeida*

*Department of Physics, Queens College, City University of New York, Flushing, NY 11367, United States of America*

*The Graduate Center of the City University of New York, New York, NY 10016, United States of America*

*Author's email address: euclides.almeida@qc.cuny.edu

## Methods

*Numerical Simulations*

The design of the device was carried out using the Lumerical FDTD package. The optical constants of silicon, gold, chromium, $SiO_2$ and $Al_2O_3$ were taken from Palik handbooks [1] and the 2D graphene conductivity model was used for graphene [2]. This model calculates the optical conductivity as composed of the intraband ($\sigma_{intra}$) and interband ($\sigma_{inter}$) conductivity terms:

$$\sigma(\omega) = \sigma_{intra}(\omega) + \sigma_{inter}(\omega)$$

The intra and interband conductivities are given by:

$$\sigma_{intra}(\omega) = \sigma_{intra}(\omega, \mu, \Gamma, T) = \frac{ie^2}{\pi\hbar^2(\omega - i2\Gamma)} \int_0^\infty \varepsilon \left(\frac{\partial f(\varepsilon)}{\partial \varepsilon} - \frac{\partial f(-\varepsilon)}{\partial \varepsilon}\right) d\varepsilon$$

$$\sigma_{inter}(\omega) = \sigma_{inter}(\omega, \mu, \Gamma, T) = \frac{ie^2(\omega - i2\Gamma)}{\pi\hbar^2} \int_0^\infty \frac{f(-\varepsilon) - f(\varepsilon)}{(\omega - i2\Gamma)^2 - 4(\varepsilon/\hbar)^2} d\varepsilon$$

Where $\mu$ is the chemical potential, $\Gamma$ is a phenomenological (graphene) damping parameter, $T$ is the temperature, e is the elementary charge, $f(\varepsilon) = \frac{1}{\exp[(\varepsilon - \mu)/k_B T] - 1}$ is the Fermi-Dirac distribution. $k_B$ is the Boltzmann constant. In the simulation of our fabricated metasurface, the phenomenological parameter was set to 0.02 eV. This model does not take into account plasmon damping by graphene phonons below 6.3 µm, and our calculations were restricted to graphene plasmons at longer wavelengths.

*Device fabrication*

A GFET S11 from Graphenea was used [3]. It is a commercially produced chip of CVD graphene transferred to a 90 nm silica layer on top of p-doped silicon. Graphenea creates the microstructure of graphene patches in contact with gold pads using photolithography. The mobility of graphene was 2472 cm2/V·s. We then created the metasurfaces on the graphene patches through an e-beam lithography process. Bilayer PMMA was spun on and exposed with an e-beam to form a deposition mask. The bottom layer is 2% solids 495K molecular weight PMMA in anisole solution (Kayaku PMMA 950K A2), and the top is 4% solids 950K molecular weight PMMA in anisole solution (Kayaku PMMA 950 A4). Both layers were individually spun at 4000 RPM with a 500 RPM/s ramp for 60 s, each spin followed by a bake at 180 C for 90 s. The resist was exposed with an Elionix 50 keV tool with a beam current of 200 pA using a shot pitch of 2 nm and 0.184 µs per shot. The deposition mask was developed in a 1:3 dilution of MIBK in IPA (60 s) followed by an IPA bath (30 s). 3 nm of Cr (adhesion layer, 0.3 Å/s) followed by 25 nm of Au (0.5 Å/s) was deposited through the mask to form the gold rods of the metasurface. A liftoff process with acetone was employed to obtain the clean metasurface seen in Figure 2(a). The GFET S11 was attached to a printed circuit board, and the gold pads were wire-bonded using an electrically conductive epoxy (EPO-TEK, H20E).

*Optical and electrical characterization*

The transmittance spectra of the device is obtained in a home-built infrared microscope using a Fourier Transform Infrared Spectrometer (Newport MIR8035) coupled to a Mercury Cadmium Telluride detector (Newport 80026 MCT). The resolution of the FTIR spectrometer was set to 8 cm$^{-1}$. A broadband SiC mid-infrared light source (Newport 80007 SiC) is focused on the sample using a gold parabolic mirror to a spot size comparable to the metasurface array. A ZnSe focusing lens collects the transmitted light and directs it on to either an infrared camera (for targeting) or to the FTIR (for spectral analysis) using a motorized flip mirror. A spectrum is also taken from a nearby section of silica/silicon without a graphene metasurface for the purposes of normalization of the spectrum to the SiC light source. A ZnSe wire grid linear polarizer (Thorlabs, WP25H-Z) is placed after the microscope to transmit infrared light polarized along the long axis of the gold rods. The thermal background with the signal blocked is also taken so that the thermal background can be removed from the data.

The chemical potential of the graphene is adjusted by applying a voltage to the silicon backgate of the device (Keithley 2400) while measuring the source-drain current at a source-drain voltage of 1 mV (Keithley 2401). Using these source-measure units, the transfer curve of the graphene device was taken. The graphene intrinsic doping slowly but noticeably shifts as a gate voltage is applied to it, which manifests as a changing source-drain current. During the experiment, a target current, which corresponds to a desired chemical potential, is selected, and the gate voltage is adjusted in real time to maintain that target current.

*Mid-infrared radiation modulation*

The frequency modulation data is obtained via the same optical setup as the transmittance measurement, but instead the light was routed directly to the MCT detector, without passing through the FTIR. Mid-infrared bandpass filters centered at 7.5 µm (Edmund Optics, #17-247) and 11.5 µm (Thorlabs, FB11500-500) were used to select different resonances of the metasurface. The gate voltage was modulated using an arbitrary waveform generator (Keithley 3390) and a voltage amplifier (Thorlabs, HVA200) to provide a sinusoidal high-voltage to the gate. The root mean squared voltage from the MCT detector was measured using a 200 kHz lock-in amplifier (Stanford Research Systems, SR 830). The MCT detector responsivity rolls off below 200 Hz.

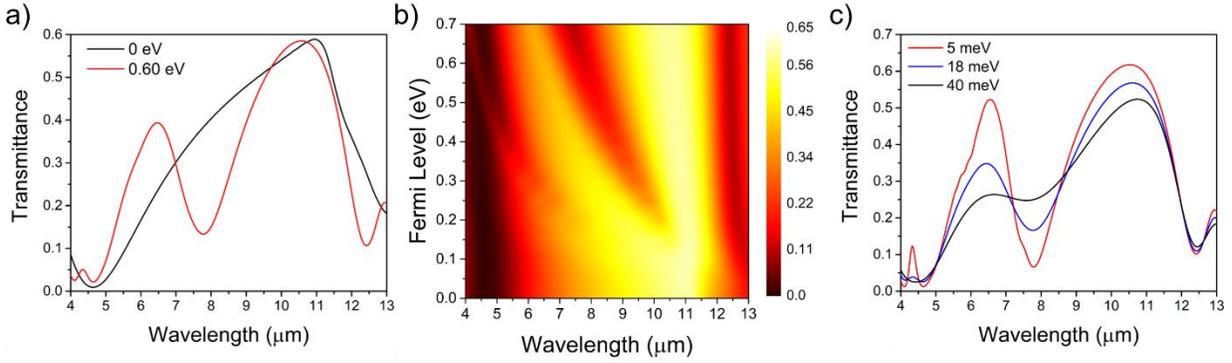

**Supplementary figure 1. Simulation of the hybrid metasurface using $Al_2O_3$ gate dielectric on silicon.** A) Calculated transmittance of the metasurface for graphene's Fermi level $E_F$ = 0 eV (black curve) and $E_F$ = 0.60 eV (red curve). The parameters of the array are (top view shown in Fig. 2(a): ax = 900 nm, ay = 60 nm, cx = 150 nm, cy = 100 nm. The height of the gold rods is h = 20 nm, and the thickness of the $Al_2O_3$ spacer is t = 15 nm. The damping parameter of graphene is 13 meV. b) Calculated transmittance of the metasurface for various $E_F$. c) Calculated for graphene's damping parameter 5 meV (red), 18 meV (blue) and 40 meV (black). In this simulation, $E_F$ = 0.6 eV.

References

[1] E. D. Palik and G. Ghosh, *Handbook of Optical Constants of Solids, Five-Volume Set* (Elsevier Science & Technology, Saint Louis, 1997).

[2] G. Hanson, Journal of applied physics **103**, 064302 (2008).

[3] *Graphenea, www.graphenea.com/collections/buy-gfet-models-for-sensing-applications/products/gfet-s11-for-sensing-applications*, 2023.